\documentclass[letterpaper, 10 pt, conference]{ieeeconf}  
\IEEEoverridecommandlockouts                              
\usepackage{cite}
\usepackage{amsmath,amssymb,amsfonts}
\usepackage{graphicx}
\usepackage{textcomp}
\usepackage{xcolor}
\usepackage[linesnumbered,ruled,vlined]{algorithm2e}
\usepackage{algpseudocode}
\usepackage{url}

\newtheorem{example}{Example}

\def\BibTeX{{\rm B\kern-.05em{\sc i\kern-.025em b}\kern-.08em
    T\kern-.1667em\lower.7ex\hbox{E}\kern-.125emX}}

\title{\LARGE \bf
A Location Validation Technique to Mitigate GPS Spoofing Attacks in IEEE 802.11p based Fleet Operator's Network of Electric Vehicles
}

\author{Ankita Samaddar$^{1}$ and Arvind Easwaran$^{2}$
\thanks{$^{1}$Ankita Samaddar is a postdoctoral researcher with the Computer Science Department, Vanderbilt University, United States. This work was done when she was affiliated with the Continental-NTU Corporate Lab, Singapore. 
        {\tt\small ankita003@e.ntu.edu.sg}}%
\thanks{$^{2}$Arvind Easwaran is a faculty with the School of Computer Science and Engineering, Nanyang Technological University, Singapore.
        {\tt\small arvinde@ntu.edu.sg}}%
}

\begin{document}

%

\maketitle

\begin{abstract}
Most vehicular applications in electric vehicles use IEEE 802.11p protocol for vehicular communications. \emph{Vehicle rebalancing} application is one such application that has been used by many car rental service providers to overcome the disparity between vehicle demand and vehicle supply at different charging stations. Vehicle rebalancing application uses the GPS location data of the vehicles periodically to determine the vehicle(s) to be moved to a different charging station for rebalancing. However, a malicious attacker residing in the network can spoof the GPS location data packets of the target vehicle(s) resulting in misinterpretation of the location of the vehicle(s). This can result in wrong rebalancing decision leading to unmet demands of the customers and under utilization of the system. To detect and prevent this attack, we propose a location tracking technique that can validate the current location of a vehicle based on its previous location and roadmaps. We used OpenStreetMap and SUMO simulator to generate the roadmap data from the roadmaps of Singapore. Extensive experiments on the generated datasets show the efficacy of our proposed technique.
\end{abstract}


\section{Introduction}\label{sec:intro}
\noindent
Rapid adoption of mobile internet technologies in vehicular networks have reshaped urban mobility providing on-demand ride services to customers. However, a large number of customers remain unserviced due to disparity between vehicle demand and vehicle supply at different charging stations. To overcome the imbalance between vehicle demand and vehicle supply, \emph{vehicle rebalancing} application has been widely adopted by many car rental service providers in recent years~\cite{marco2012,brar2020,miao2019}. In a vehicle fleet management framework, the vehicle rebalancing application predicts the future demands of the vehicles at each charging station and redistributes the 
vehicles to the stations with higher demands so that the waiting time of the customers is reduced and the fleet utilization is maximized. 

\begin{figure}[t]
    \centering
    \includegraphics[height=4cm]{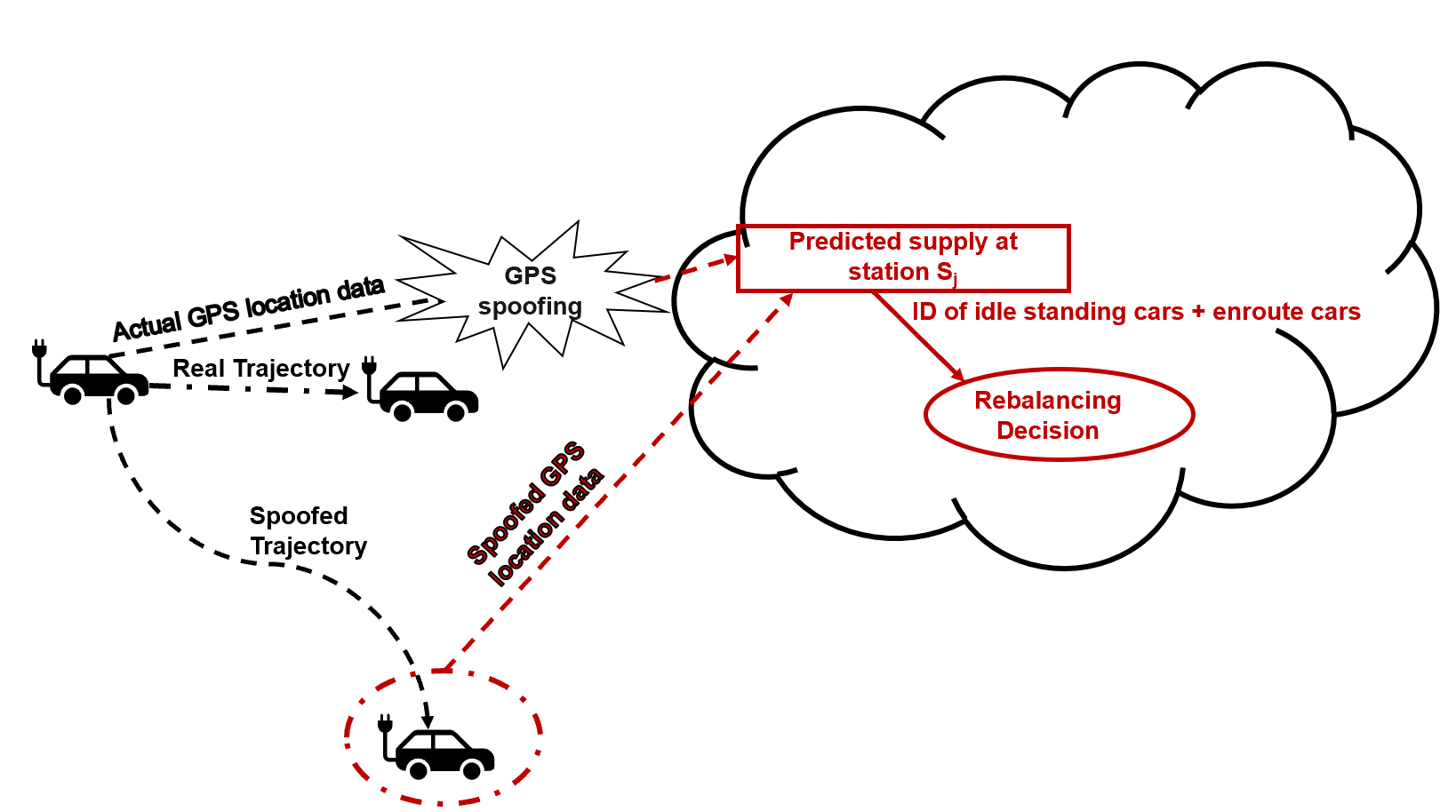}
    \caption{GPS spoofing attack in vehicle rebalancing application}
    \label{fig:gpsspoof}
\end{figure}

Among the existing network protocols, the IEEE 802.11p protocol is most popularly used for vehicular communications~\cite{wbss}. 
Each vehicle in the fleet uses IEEE 802.11p to transmit its GPS location information periodically to the cloud server or the backend server. The vehicle rebalancing application uses this GPS location information of the vehicles to predict the vehicle supply at each charging station, and select the appropriate vehicle(s) to rebalance. However, vehicular networks such as IEEE 802.11p networks are often exposed to different types of cyber-attacks such as spoofing, eavesdropping, man-in-the-middle attacks, etc~\cite{cybersecurity2020,cyberattack2021}. A malicious attacker residing in close vicinity of a target vehicle in the fleet can spoof the GPS location information of the vehicle while it is being delivered to the backend server. Spoofing of GPS location information of the target vehicle(s) results in misinterpretation of the location information of the vehicle.
This, in turn, can result in wrong rebalancing decision, creating long waiting time of the customers, eventually leading to unmet demands and under utilization of the fleet. Fig.~\ref{fig:gpsspoof} shows a rough sketch of GPS spoofing attack in vehicle rebalancing application. 

Although GPS spoofing attack has been well studied in the literature, all of these studies consider GPS spoofing attack while GNSS signal is being received by the onboard units~\cite{gpsspoof2012,gpsspoofing2021}. However, in this work, we consider GPS spoofing attack when the GPS location information is transmitted from the onboard units of the vehicles to the backend server using IEEE 802.11p. As a countermeasure against this attack, we propose a location tracking technique that can validate the current location of the vehicle based on its previous location and the roadmap data. Further, we use a Hashed MAC authentication protocol to authenticate the messages if an attack is detected at the backend server~\cite{sha512}. To the best of our knowledge, this is the first work in the literature that addresses GPS spoofing attack in vehicle rebalancing application and a countermeasure to mitigate the attack. 

Thus, the main contributions of this work are:

\noindent
1. We propose a GPS spoofing attack in vehicle rebalancing application for a fleet of electric vehicles.

\noindent
2. We propose a location tracking technique to validate the current location of a vehicle based on its previous location and roadmap data. We use a Hashed MAC authentication protocol to prevent the attack.  

\noindent
3. We generated roadmaps of Singapore using OpenStreetMap~\cite{openstreetmap} and SUMO simulator~\cite{SUMO2018}, and ran our experiments on the generated datasets. We used an error function to evaluate our proposed technique. Experiments on the generated datasets show that the error value always remains below $1.0$ in absence of GPS spoofing attack under all conditions. Further, on launching GPS spoofing attack, the error value immediately exceeds $1$ detecting the attack. 

\section{Background}\label{sec:background}
\noindent
In this section, we presented vehicle rebalancing application and IEEE 802.11p protocol in detail.

\vspace{0.5em}\noindent
\textbf{Vehicle Rebalancing Application : } An $A$-to-$B$ electric car rental business model provides car rental services to customers who can arrive at a charging station, rent an electric car and go for a ride. After completion of the ride, the customer can drop the car at any charging station. 
Due to disparity in the popularity of the stations, variation in traffic at different times of the day, weather conditions, public holidays, etc. the station based $A$-to-$B$ car rental systems suffer from spatial-temporal supply-demand imbalances, \emph{i.e.}, at some point in time a situation may arise when there are no cars at some stations despite having high demand, while, there are idle standing cars at other stations. This results in under utilization of the fleet at some stations while creating high waiting times for the customers at other stations, eventually leading to unmet demands. 

To overcome this mismatch in the demand versus supply of vehicles, \textit{vehicle rebalancing} application has been adopted by many car rental service providers 
to rebalance the supply of vehicles across the stations so that the utilization of the fleet gets maximized. For example, if the predicted demand of vehicles at station $A$ over a time window $t$ surpluses the predicted supply of vehicles at station $A$ over that time window, then the vehicle rebalancing application selects an idle standing vehicle from the neighbouring stations of station $A$ and move that vehicle to station $A$ so that it can maximize the supply at $A$ and meet the predicted demand at $A$. However, in absence of any dedicated driver in a car-rental setting, the vehicles cannot be moved from one station to another on its own. Instead, a dynamic pricing scheme is associated with the vehicle rebalancing decision that influences the choice of the pick-up and drop-off station of the customers. The dynamic pricing scheme dynamically generates the ride fares to increase the utilization of the fleet. 

\vspace{0.5em}\noindent \textbf{IEEE 802.11p :} A majority of applications in electric vehicles that are associated with road safety message transmissions between the vehicles and the roadside units or the backend server, uses IEEE 802.11p protocol or the Wireless Access in Vehicular Environments (WAVE)~\cite{a2020arena}. 
IEEE 802.11p operates at $5.9$GHz frequency 
over a bandwidth of $75$MHz and uses carrier sense multiple access (CSMA) based communications. The entire bandwidth is divided into six service channels (SCH), one control channel (CCH) and one reserved channel. The CCH is used to broadcast control messages and the SCHs are used for application specific communications. The SCH requires the setup of a WAVE Basic Service Set (WBSS) prior to its usage~\cite{wbss}. A WBSS in IEEE 802.11p consists of a group of wireless devices over a specific region that share the same physical layer medium access characteristics such as radio frequency, modulation scheme, security settings, etc. so that they are wirelessly networked~\cite{wbss}. As the vehicles move from a region to another, they pass from one WBSS to another. Each WBSS has a unique ID. In WAVE mode, data packet transmissions are allowed within an independent WBSS without any active scanning, association or authentication. The backend server periodically broadcasts a WAVE Service Advertisement (WSA) message on the CCH without any acknowledgement on their successful reception. The WSA contains all the information to identify the WAVE application, the network parameters necessary to join a WBSS such as the WBSS ID, the SCH used by the WBSS, the timing information for synchronization purposes, etc. A vehicle waiting to join a WBSS first monitors the WSAs on the CCH, learns the operational parameters and then joins the WBSS by switching to the dedicated SCH used by the WBSS. After successfully joining the network, the vehicles belonging to a specific WBSS contend among themselves to send the GPS location data by CSMA based on the SCH access pattern.

\section{Related Works}\label{sec:related}
\noindent
Vehicular networks are susceptible to a multitude of cyber-attacks. GPS spoofing attack is a type of man-in-the-middle attack where the attacker compromises the system by spoofing the location information of the target vehicle~\cite{gpsspoofing2021}. Kamal \emph{et al.} and Zhenghao \emph{et al.} illustrated spoofing of GPS signals as they are transmitted by GNSS and received by the onboard units~\cite{gpsspoof2012,gpsspoofing2021}. Ledvina \emph{et al.} proposed to adopt an in-line RF device with a GPS antenna and a legacy civil GPS receiver at the physical layer to mitigate GPS spoofing attack when GNSS signal is received~\cite{inlinespoofing2010}. Stephen \emph{et al.} presented vulnerabilities across multiple attack surfaces in automotive systems~\cite{comprehensive2011}. Jiang \emph{et al.} proposed ``DeepPOSE", that analyzes the GNSS signals received from the satellite and detects GPS spoofing attack using recurrent neural network~\cite{Deeppose2022}. Mykytyn \emph{et al.} proposed a GPS spoofing attack detection mechanism in UAV swarm networks based on the distance between any two swarm members~\cite{gpsspoofing2023}. However, none of these works consider GPS spoofing attack when the GPS location data is transmitted to the backend server in a fleet via the onboard units using IEEE 802.11p. Demissie \emph{et al.} presented a truck movement model using origin-destination flows, destination choices and GPS data in truck movement~\cite{demissie2022}. However, this solution is not applicable to our setting to detect GPS spoofing attack. Thus, in this work, we propose a countermeasure to mitigate GPS spoofing attack in electric vehicles that validates the current GPS location of a vehicle in a fleet based on its previous GPS location and the roadmap data. To the best of our knowledge, this is the first work that proposes GPS spoofing attack in the context of vehicle rebalancing application and a location validation technique as a countermeasure to mitigate the attack.

  
  
   


\section{System Model}\label{sec:system}
\begin{figure}
\centering
		\includegraphics[height=4cm]{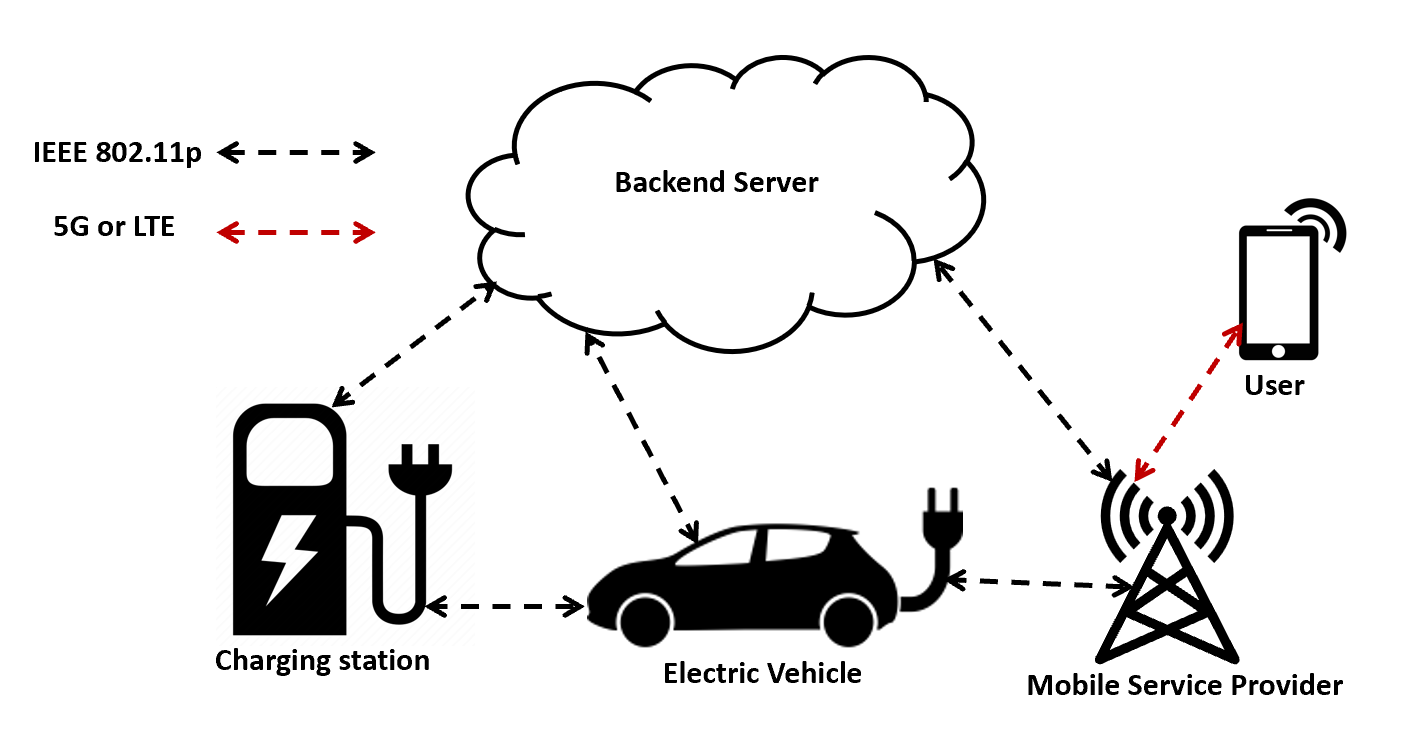}
\caption{Data transmissions among different components of a fleet}	
\label{fig:datacomm}
\end{figure}

\noindent
We consider a vehicular network consisting of a fleet of $n$ electric vehicles, $V = \{V_1,V_2,\ldots,V_n\}$, a set of $k$ charging stations, $S = \{S_1,S_2,\ldots,S_k\}$, that are spatially distributed across a region/city, a backend server or a cloud server $B$, and users using the car rental services provided by the fleet operator. A vehicle is charged at a charging station whenever its charge falls below a certain threshold. On completion of a ride, the user parks the vehicle in the nearby charging station based on parking availability. The backend server stores information about the vehicles in the fleet, the charging stations, the users in the system and the roadmaps of the city and performs different 
computations associated with the fleet. Fig.~\ref{fig:datacomm} shows the wireless data transmissions in the fleet operator's network. The users communicate with the fleet operator via 5G or LTE. All other vehicular data transmissions in the network use IEEE 802.11p.

We represent the roadmaps of a particular region/city as a graph $G=\{J,R\}$, where $J$ denotes the set of junctions or vertices connecting two or more roads and $R$ denotes the set of roads. Each road $r_i \in R$ is associated with a unique ID, length and a set of GPS co-ordinates along the road. Each junction $j_i \in J$ is associated with a unique ID and a GPS co-ordinate. Each vehicle $V_i \in V$ sends a location data packet periodically to the backend server $B$ through the onboard unit to inform its current location. A location data packet of a specific vehicle consists of the vehicle ID, the location of the vehicle (latitude, longitude, bearing angle) and the timestamp at which the vehicle sends the data packet to the backend server. The backend server $B$ maintains a table to store the most recent GPS location data (latitude, longitude, bearing angle) corresponding to each vehicle in the fleet along with the timestamp at which it was sent.

\vspace{0.5em}\noindent
\textbf{Steps of Vehicle Rebalancing application:} The vehicle rebalancing application consists of the following steps.

\noindent
\textbf{Step 1 : Predict the vehicle demand at each station.} A vehicle demand prediction model runs at the beackend server $B$ every $\Delta t$ time units to predict the vehicle demand $d^{j}_{t,t+\Delta t}$ at station $S_j$ over a time window [$t$,$t+\Delta t$]. 

\noindent
\textbf{Step 2 : Predict the vehicle supply at each charging station.} A vehicle supply prediction model runs at $B$ every $\Delta t$ time units to predict the supply of vehicles $s^j_{t,t+\Delta t}$ at station $S_j$ over a time window [$t$,$t+\Delta t$]. The supply of vehicles at station $S_j$ over time window [$t$,$t+\Delta t$] is determined by the number of idle standing cars at $S_j$ and the number of enroute cars towards $S_j$. The backend server keeps track of the total number of idle standing cars at each station. Whenever a car parks at a station or leaves a station, this information is updated at the backend server accordingly. 
The backend server uses the \emph{GPS location data} of the vehicles periodically to predict the number of enroute cars at $S_j$.  

\noindent
\textbf{Step 3 : Rebalance the vehicles from the neighbouring stations.} If the predicted demand at station $S_j$ is greater than the predicted supply at $S_j$ over time horizon [$t$,$t+\Delta t$], rebalance the cars from the neighbouring stations to station $S_j$. To rebalance the cars from the neighbouring stations, a dynamic pricing scheme is associated with the vehicle rebalancing application that determines the ride fares of the vehicles that are to be rebalanced. 

\noindent
\textbf{Step 4 :} Repeat Step~1 to Step~3 every $\Delta t$ time units. Generally, this $\Delta t$ is variable and depends on various parameters such as time of the day, traffic condition, weather condition, popularity of the charging stations, etc.

\begin{figure}
\centering
		\includegraphics[width=8cm]{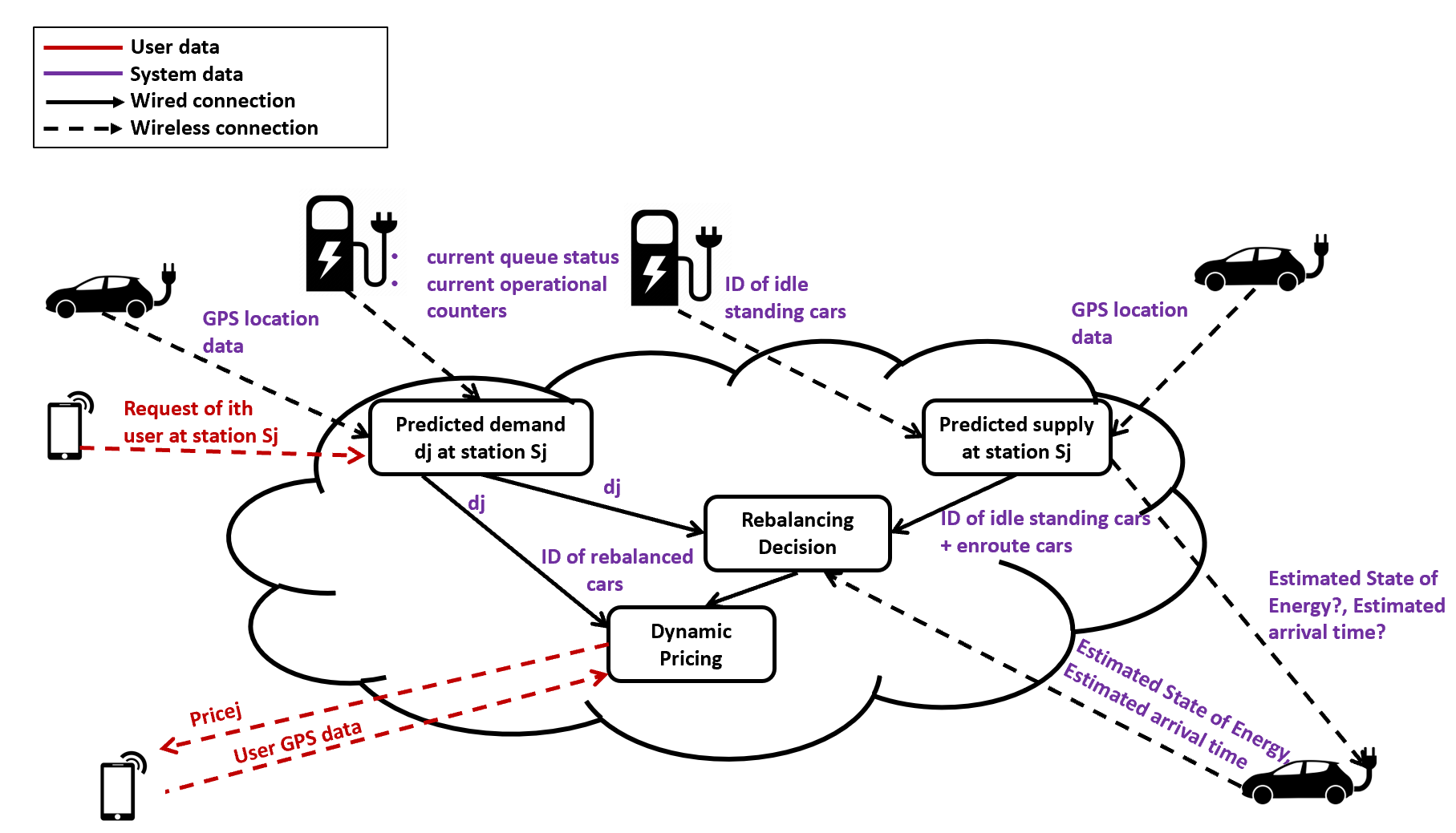}
\caption{Data Flow Diagram of Vehicle Rebalancing Application}	
\label{fig:dfd}
\end{figure}

Fig.~\ref{fig:dfd} shows the data flow diagram of the vehicle rebalancing application. The wired/wireless transmissions are marked with solid/dotted arrows. The user/system data are marked with red/purple colors.



\section{Attack Model}\label{sec:attack}
\noindent
We consider a stealthy attacker whose main objective is to spoof or alter the GPS location data of the target vehicle(s) while the location data packet is being transmitted from the target vehicle(s) to the backend server. Altering the GPS location data of the vehicle results in transmission of wrong GPS location information to the backend server. This leads to wrong supply prediction at station $S_j$ and eventually results in wrong rebalancing decision at the backend server. 

To launch GPS spoofing attack, our attacker has the following assumptions. 

\noindent
1. The attacker is a malicious device equipped with a directional antenna and residing in close proximity of the target vehicle. It can even be a vehicle in the fleet operator's network with an onboard antenna behaving maliciously~\cite{directional}.
    
\noindent
2. The attacker lies within the same region, \emph{i.e.}, within the same WBSS of the target vehicle and can listen to the WSA message that is broadcasted periodically from the backend server. Thus, the attacker knows the SCH access pattern within that WBSS. 

\noindent
3. The attacker is equipped with a timer to determine the time difference between two successive GPS data transmissions from the target vehicle to the backend server.

Based on the above assumptions, the attacker has the following capabilities.

\noindent
\emph{Capability 1: The attacker can use its directional antenna to eavesdrop the transmissions between the target vehicle and the backend server.}

\noindent
\emph{Capability 2: Since the attacker and the target vehicle resides in the same WBSS, it can listen to the WSA that are broadcasted periodically from the backend server to all the vehicles within the same WBSS. It can also estimate the SCH access pattern by analyzing the broadcasted WSA.}

\noindent
\emph{Capability 3: The attacker can use its timer to estimate the time interval between two successive GPS transmissions from the target vehicle to the backend server through the SCH.}

\noindent \textbf{Description of the attack:} Spoofing of GPS location data occurs in two phases: 

\textbf{Phase~1: Determine the transmissions from the target vehicle:} The attacker determines the SCH access pattern based on the WSA received from the backend server. It then eavesdrops the SCH based on its access pattern to determine the time slots in which the target vehicle transmits GPS location data to the backend server. While eavesdropping, the attacker uses its timer to estimate the time interval between two successive GPS location data transmissions from the the target vehicle to the backend server.

\textbf{Phase~2: Spoofing GPS location data:} On identifying the SCH access pattern and estimating the time interval between any two successive GPS location data transmissions from the target vehicle to the backend server, the attacker can predict the time slots and the SCH to be used by the target vehicle to transmit successive GPS location data packets. Thereafter, the attacker eavesdrops the SCH in the predicted time slots and extracts important data fields such as the Vehicle ID, latitude, longitude, or timestamp. Later, the attacker can interfere subsequent transmissions in the predicted slots, alter GPS information such as latitude or longitude information, and can resend the data packets to the backend server.
Additionally, the attacker can alter the timestamp information in the eavesdropped data packets and can replay the same data packet to the backend server at a later point in time.

\begin{figure}
    \centering
    \includegraphics[width=8cm]{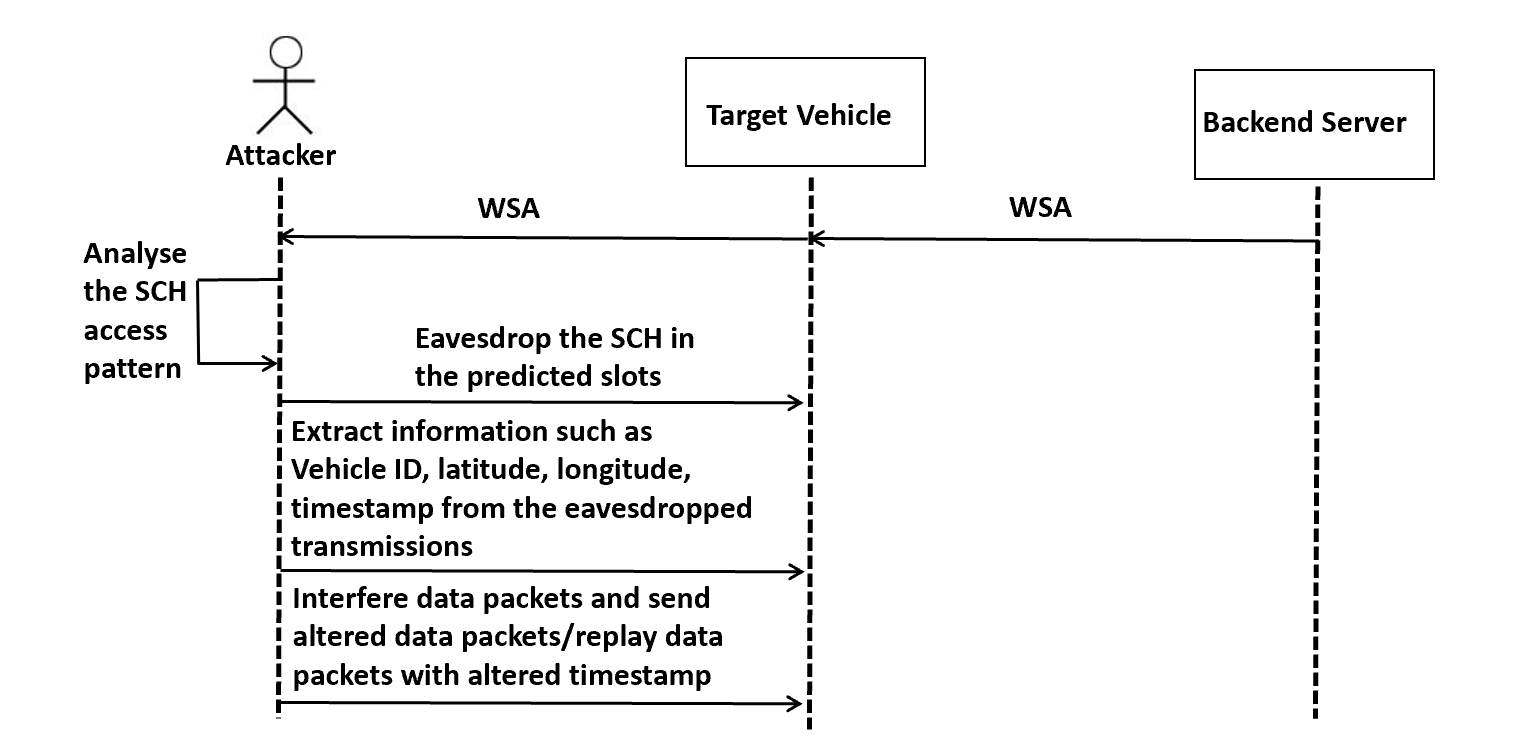}
    \caption{Sequence diagram illustrating the interactions in sequence to launch GPS spoofing attack}
    \label{fig:seqdiag}
\end{figure}

Fig.~\ref{fig:seqdiag} shows the interactions in sequence to launch GPS spoofing attack.

\noindent \textbf{Attack Consequence :} The GPS location information plays a vital role in decision making for many vehicular applications such as the vehicle rebalancing application. In this application, the supply prediction at any station $S_j$ is determined by the GPS location data sent by the vehicles to the backend server (Step~2 of vehicle rebalancing application in Section~\ref{sec:system}). GPS spoofing attack leads to transmission of altered GPS location data to the backend server resulting in wrong rebalancing decision. 

We present an example to show the attack consequences.

\begin{example}
    \emph{Consider a GPS location data packet with the following information; \textbf{Vehicle ID: 10010, Latitude: 1.302; Longitude: 84.24; Bearing Angle: 30; Timestamp: 09/05/2024:11:05:24}. On spoofing the data packet and altering 1 bit in the longitude information, the attacker sends an altered data packet; \textbf{Vehicle ID: 10010, Latitude: 1.302; \textcolor{red}{Longitude: 80.24;} Bearing Angle: 30; Timestamp: 09/05/2024:11:05:24}. Due to alteration of the longitude information, the backend server will calculate a different estimated arrival time for this vehicle. It may happen that due to spoofed location, a vehicle may appear to be nearer to station $S_j$ and considered in the rebalancing decision although its actual position is far away from $S_j$. Or, it may happen that due to spoofed location, it may appear to be far away from station $S_j$ and not considered in the rebalancing decision although its actual position is very near to $S_j$. Any of these can lead to long waiting times of the customers, causing under utilization of the fleet resulting in unmet demands. By targeting a significant percentage of the vehicles in the fleet, this attack can result in a large number of unmet demands thereby incurring loss to the fleet operator.}
\end{example}

\section{Countermeasure to the Attack}\label{sec:countermeasure}
\noindent
As a countermeasure against GPS spoofing attack, we propose an online attack detection and prevention technique at the backend server. Our mitigation technique uses roadmaps to validate the current location of each car in the fleet based on its previous location. If our detection technique detects an error in the current location of the car, then the backend server requests the car to authenticate a few subsequent transmissions to prevent GPS data spoofing. 

Our countermeasure consists of three phases. 
\begin{enumerate}
    \item An \emph{offline Pre-Processing Phase} that runs at the backend server during network initialization
    \item An \emph{online Attack Detection Phase} that runs at the backend server 
    \item An \emph{online Attack Prevention Phase} that runs at the backend server only if an attack is detected
\end{enumerate}

\vspace{0.5em}\noindent
\textbf{Offline Pre-Processing Phase:} Algorithm~\ref{algo:preprocess} shows the pre-processing phase. In the pre-processing phase, the details of each vehicle (vehicle ID, key) in the fleet (lines~1-2) and each charging station (station ID, GPS location) are stored in the backend server (lines~3-4). For each junction in the roadmap of the region/city, the unique junction ID and GPS data (latitude, longitude) are stored in $Jn\_List$ and list $L$ (lines~6-8). For each road $r_i$ in the roadmap, the road id ($r_i.id$), road length ($r_i.len$), maximum and minimum GPS coordinates on that road  ($r_i.min$ and $r_i.max$) and the maximum allowable speed of the road ($r_i.sp$) are stored (line~10). Additionally, the GPS data along each road $r_i$ at $1$meter apart are stored in $r_i.T$ (lines~9-12). For each junction $j_i$, the set of roads converging at $j_i$ are stored in the adjacency list $L[j_i]$ (line~14-15). A dictionary $Rd\_Jn$ is created to store the junction id associated with any two roads in the roadmap (lines~17-18).  

\begin{algorithm}[t]
    \small
    \For{each $V_i \in V$}
    {
        Assign $V_i.id$, $V_i.key$ to $V_i$\;
    }
    \For{each $S_i \in S$}
    {
        Assign $S_i.id$, $S_i.loc$ to $S_i$\;
    }
    $L =$ [];\tcp{List storing junction IDs and roads originating at junction}
    \For{each $j_i \in G.J$}
    {   
        \tcp{Add junction details to $Jn\_List$,$L$}
        Assign $j_i.id$, $j_i.loc$ to $j_i$\;
        Add $j_i$ to $Jn\_List$ and $L$\;
    }
    \tcp{Assign road data to $Rd\_List$}
    \For{each $r_i \in G.R$}
    {
        Add $r_i.id$, $r_i.len$, $r_i.min$, $r_i.max$, $r_i.sp$ to $r_i$\;
        $r_i.T =$ [];\tcp{List storing road data}
        Add GPS locations (latitude, longitude, bearing angle) on the road $r_i$ at $1$meter apart to $r_i.T$\;
        Add $r_i$ to $Rd\_List$\;
    }
    
    \For{each $j_i \in G.J$}
    {
        $L[j_i] =$ set of roads that begins/ends in $j_i$\;
        $Rd\_Jn = \{\}$;\tcp{dictionary of roads and associated junctions}
        \For{each $r_j$ and $r_k$ in $L~|~r_j \neq r_k$}
        {
            $Rd\_Jn[r_j\_r_k]$ = $Rd\_Jn[r_k\_r_j]$ = $j_i$\; 
        }
    }
    \caption{\textit{PreProcess()}}
    \label{algo:preprocess}
\end{algorithm}

\begin{figure}
    \centering
    \includegraphics[height=3.5cm]{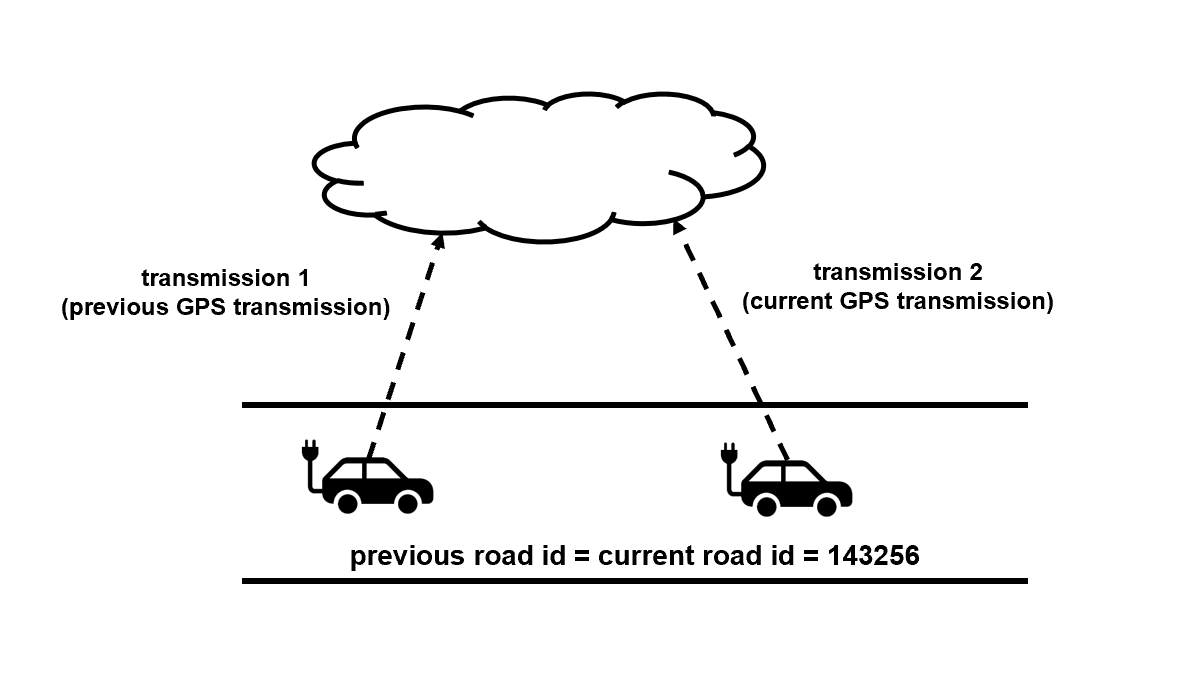}
    \caption{Previous and current locations of the vehicle are on the same road.}
    \label{fig:case1}
    \centering
    \includegraphics[height=3.5cm]{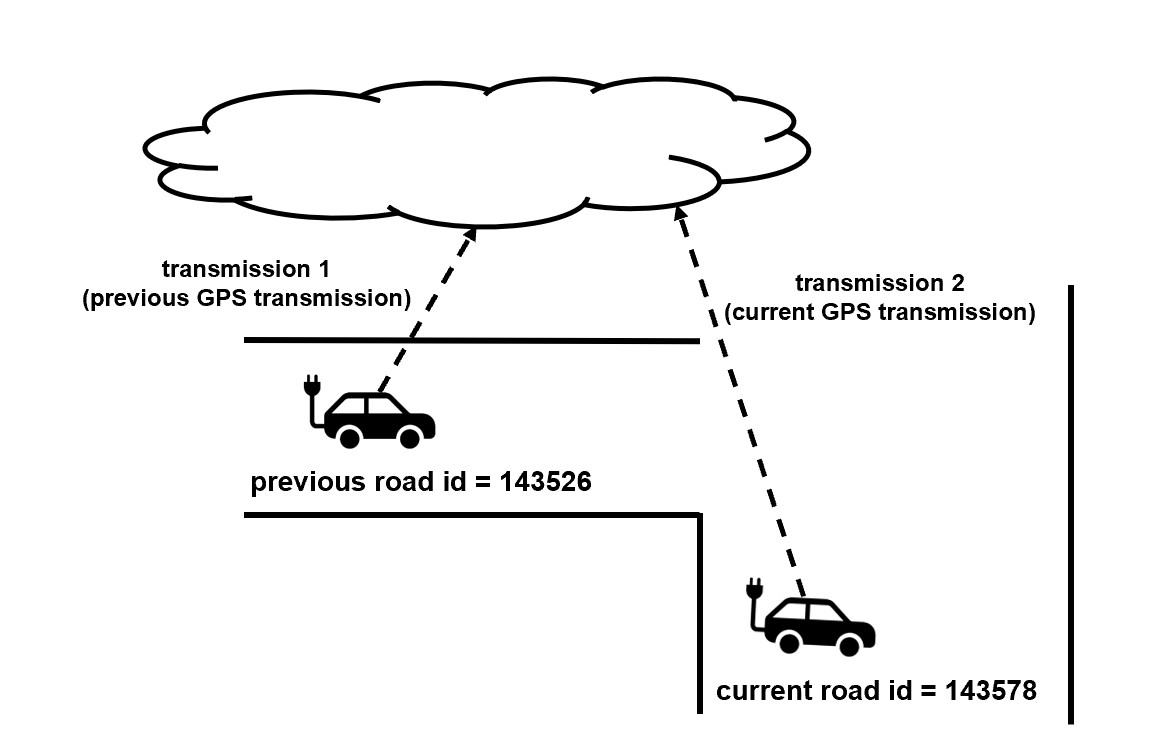}
    \caption{Previous and current locations of the vehicle are on two different roads.}
    \label{fig:case2}
    \centering
    \includegraphics[height=3.5cm]{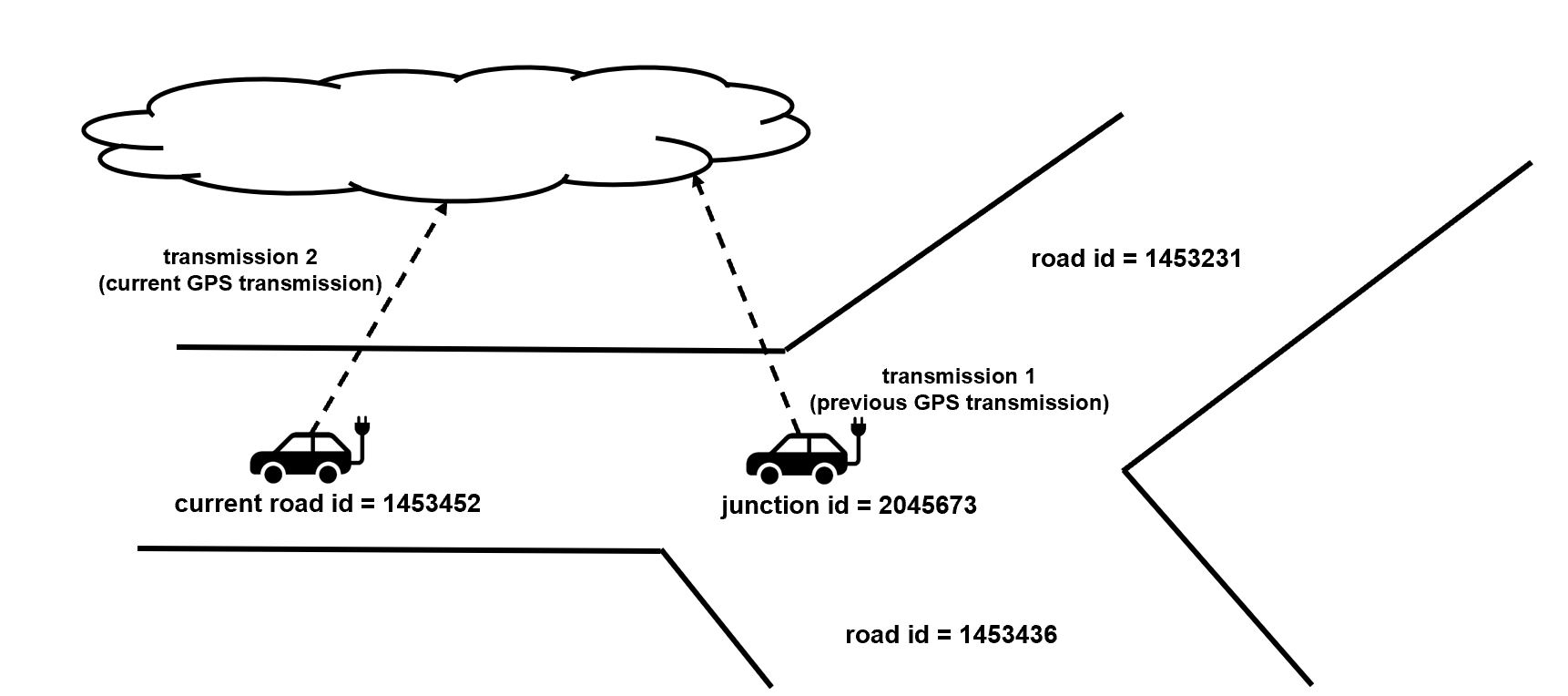}
    \caption{Previous location of the vehicle is in a junction and current location of the vehicle is on a road.}
    \label{fig:case3}
    \centering
    \includegraphics[height=3.5cm]{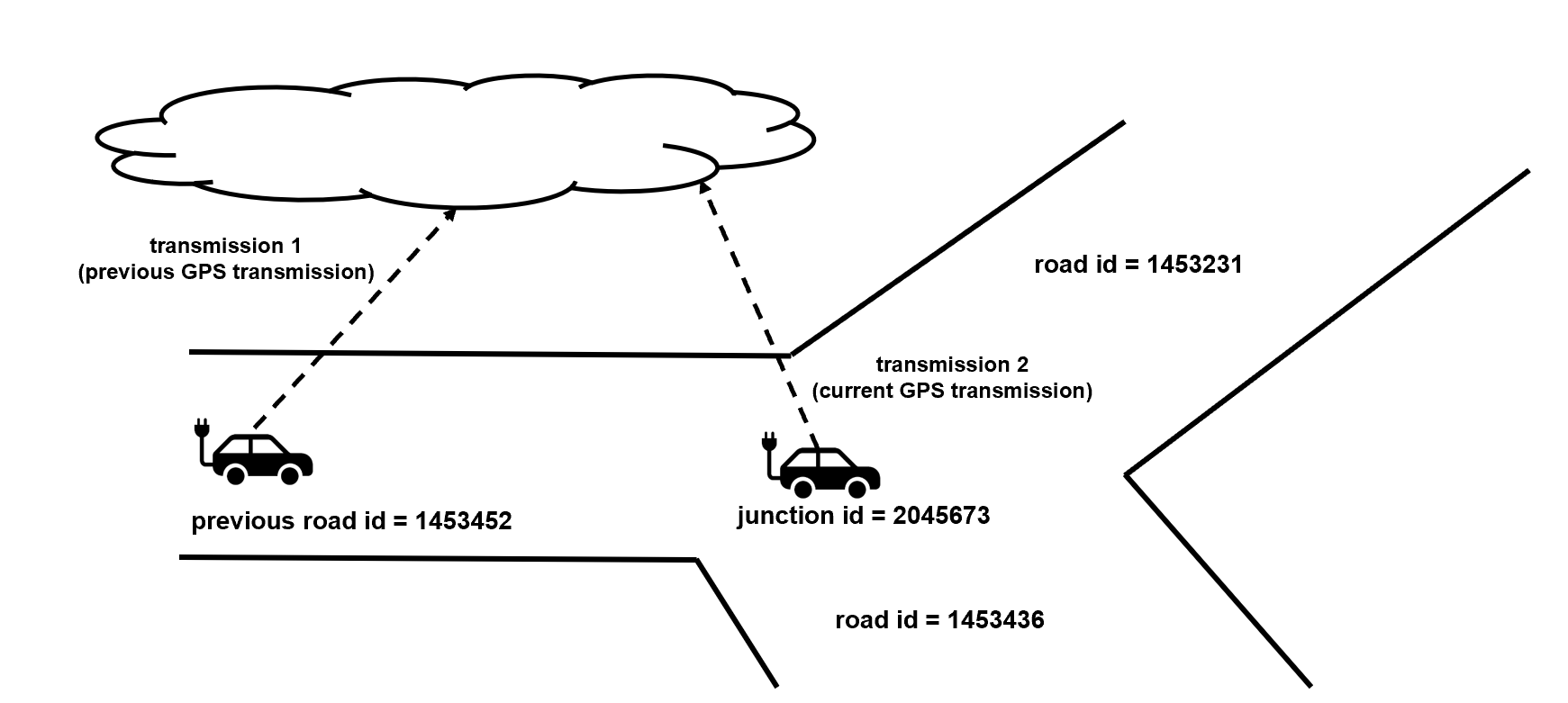}
    \caption{Previous location of the vehicle is on a road and current location of the vehicle is in a junction.}
    \label{fig:case4}
\end{figure}

\vspace{0.5em} \noindent
\textbf{Online Attack Detection Phase:} Algorithm~\ref{algo:attackdetect} presents the steps to detect GPS spoofing attack on receiving the data packet $P$ (say). From each received data packet $P$, the vehicle ID, current GPS location and timestamp is extracted (lines~4-6). The $CheckPos()$ function (line~7) validates the current vehicle position based on its previous location and roadmap. In this function, the list of roads, $l$, is generated so that the current location of the vehicle falls within the minimum and maximum GPS coordinates on that road. If $l$ is not an  empty list, then the GPS coordinate on the road that matches maximally with the current GPS location is chosen (lines~18-19). Note that, $D(x,y)$ denotes the distance between GPS coordinates $x$ and $y$. Four cases can occur based on the current and previous location of the vehicle on the road. We explain each of these cases with a figure.

\noindent
\emph{Case~1: The current and previous locations of the vehicle are on the same road (lines~20-22).} Refer to Fig.~\ref{fig:case1}. In this case, the distance between two GPS coordinates on the same road that maximally matches with the previous and current GPS locations provide the distance travelled by the vehicle between two successive GPS transmissions.

\noindent
\emph{Case~2: The current and previous locations of the vehicle are on different roads (lines~23-27).} Refer to Fig.~\ref{fig:case2}. In this case, the two GPS coordinates on the two different roads that match maximally with the previous and current locations respectively are chosen. Additionally, the $Rd\_Jn$ dictionary is searched to check if the two roads are connected by a junction. If they are connected, then the distance between the junction and the two different GPS coordinates on the two roads provide the total distance travelled between two successive transmissions.

\noindent
\emph{Case~3: The current location of the vehicle is on a road and previous location of the vehicle is in a junction (lines~28-31).} Refer to Fig.~\ref{fig:case3}. In this case, the junction with which the previous GPS location matches maximally and which has a connectivity with the current road is chosen (line~27). The distance between the junction and the current GPS location provides the distance travelled between two successive transmissions.

\noindent
\emph{Case~4: The current location of the vehicle is in a junction and the previous location of the vehicle is on a road (lines~32-37).} Refer to Fig.~\ref{fig:case4}. In this case, the junction with which the current GPS location matches maximally and which has a connectivity with the previous road is chosen (line~34). The distance between the junction and the previous GPS location provides the distance travelled between two successive transmissions.

\vspace{0.5em}\noindent
\textbf{Online Attack Prevention Phase: } The online $AttackPrevent()$ algorithm (Algorithm~3) is called if $CheckPos()$ function cannot validate the current position of the vehicle or if the distance between two successive GPS transmissions exceeds the maximum distance that the vehicle can cover in that duration, \emph{i.e.,} $\frac{d}{Max\_Dist} > 1$ (line~8 of Algorithm~2). The maximum distance ($Max\_Dist$) that a vehicle can cover between two GPS transmissions is calculated by the product of the maximum allowable speed of the road, $r_i.sp$, and the time between the two transmissions. In the attack prevention phase, $B$ sends an alert message to the victim vehicle. Thereafter, $B$ uses Hashed MAC (HMAC) authentication mechanism to authenticate subsequent $t$ GPS transmissions. The HMAC authentication protocol uses the unique key, $V_i.key$, (line~2 of Algorithm~1) to authenticate the transmission associated with vehicle $V_i$. Each vehicle in the fleet has its own unique key stored securely in the vehicle during system initialization. Additionally, the backend server securely maintains a table to store all the IDs of the vehicles in the fleet and the unique keys associated with them. We assume that this unique key is secure and not accessible to the attacker. However, we do not propose to authenticate all the GPS data transmissions in the network. This is because, authenticating all the data packets will incur delays in the network, which in turn, will make the solution non scalable.

\begin{algorithm}[t]
    \small
    Assign initial location to $prv$, road ID to $r^{prv}$, timestamp to $t^{prv}$\;
    \While{(1)}
    {
       \For{each received GPS location data packet $P$}
        {
            $V.id = P\rightarrow id$\;
            $cur = P\rightarrow (lat,lon,angle)$\;
            $t^{cur} = P \rightarrow timestamp$\; 
            $d = CheckPos(V.id,prv,cur)$\;
            \If{$\frac{d}{Max\_dist} > 1)$}
            {
                $AttackPrevent(V.id,V.key)$\;
            }
            $prv = cur$\; 
            $r^{prv} = r^{cur}$\;
            $t^{prv} = t^{cur}$\; 
        }
    }
    \SetKwFunction{Fun}{\textbf{Function:}$CheckPos$}
    \Fun{$V.id,prv,cur$}
    {
        
        $l =$ []\;
        \For{$r_i \in G.R~|~cur$ lies within [$r_i.min,r_i.max$]}
        {
            Add $r_i$ to $l$\;
        }
        \If{$len(l) > 0$}
        {
            $m_1 = min(D(r_i.T[k],cur))~|~k \in [1,len(r_i.T)]~\&\&~r_i \in l$\;
            $r^{cur} = r_i~|~m_1 \in r_i.T$\;
            \If{$r^{cur} == r^{prv}$}
            {
                $m_2 = min(D(r^{prv}.T[k],prv))~|~k \in [1,len(r^{prv}.T)]\}$\;
                return $D(m_1,m_2)$\;
            }
            \ElseIf{$r^{prv} \neq -1~\&\&~r^{cur} \neq r^{prv}$}
            {
                $m_2 = min(D(r^{prv}.T[k],prv))~|~k \in [1,len(r^{prv}.T)]\}$\;
                \If{$(Rd\_Jn[r^{cur}\_r^{prv}] \neq null)$ ~$or$~$(Rd\_Jn[r^{prv}\_r^{cur}] \neq null)$}
                {
                    $j_k = Rd\_Jn[r^{cur}\_r^{prv}]$\;
                    return $D(m_2,j_k) + D(m_1,j_k)$\;
                }
            }
            \ElseIf{$r^{prv} == -1$}
            {                
                $jn = j_k~|~min(D(j_k.loc,prv))~\&\&~r^{cur} \in L[j_k]$\;
                \If{$jn \neq null$}
                {
                    return $D(m_1,jn.loc)$\;
                }
            }
        }
        \Else
        {
            $m_1 = min(D(r^{prv}.T[k],prv))~|~k \in [1,len(r^{prv}.T)]\}$\;
            $jn = j_k~|~min(D(j_k.loc,cur))~\&\&~r^{prv} \in L[j_k]$\;
            \If{$jn \neq null$}
            {
                $r^{cur} = -1$\;
                return $D(m_1,jn.loc)$\;
            }            
        } 
    }
    return $100$;\tcp{Invalid location data}
    \caption{\textit{AttackDetect(GPS location data)}}
    \label{algo:attackdetect}
\end{algorithm}
\begin{algorithm}
    \small
    Transmit alert from $B$ to vehicle with vehicle ID $V.id$\;
    \For{next $t$ transmissions associated with $V.id$}
    {
        $HMAC\_Authenticate$($V.id$,$V.key$)\;
    }
    \caption{\textit{AttackPrevent($V.id$,$V.key$)}}
    \label{algo:attackprevent}
\end{algorithm}
\section{Experiments and Evaluation}\label{sec:experiment}
\begin{figure}
    \centering
    \includegraphics[height=5cm]{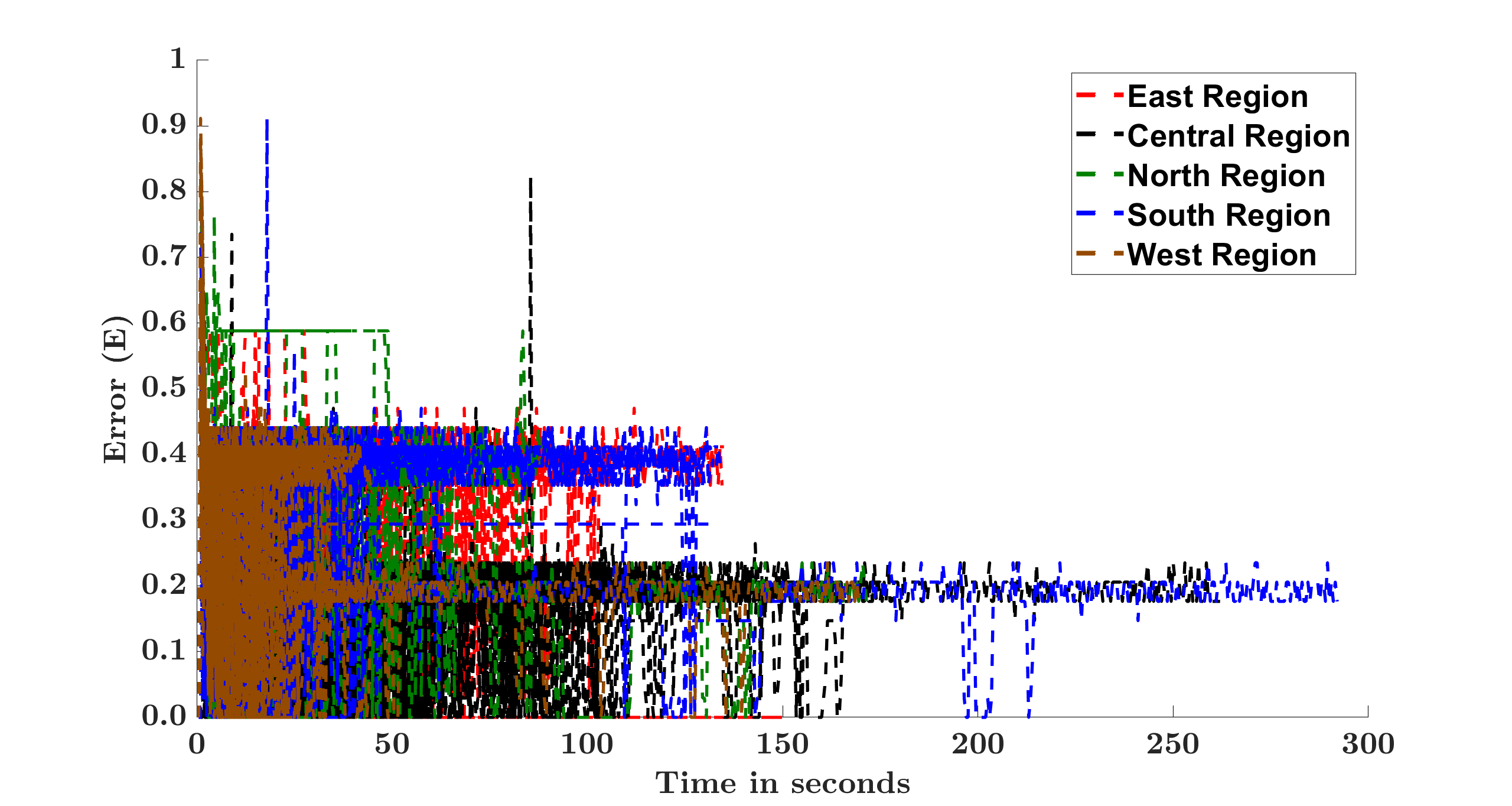}
    \caption{Error function ($E$) on $380$ trips across Singapore over 10 minutes}
    \label{fig:error}
    \centering
    \includegraphics[height=5cm]{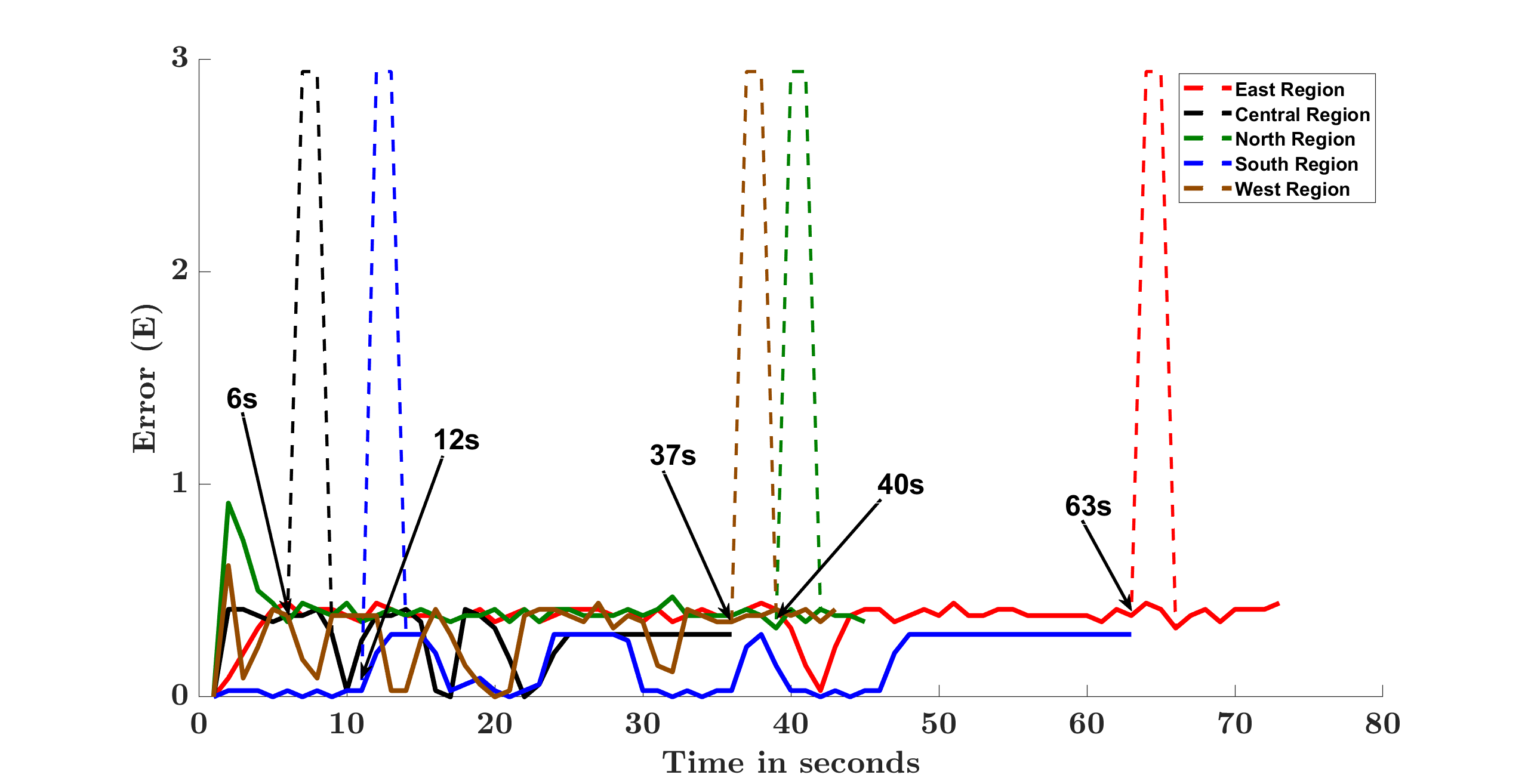}
    \caption{Random trajectories in different regions of Singapore without GPS spoofing attack (solid lines) and with GPS spoofing attack (dashed lines)}
    \label{fig:attackseries}
\end{figure}

\noindent
In our experiments, we detected GPS spoofing attack using an error function $E$ (line~8 of Algorithm~\ref{algo:attackdetect})  which is denoted by
\begin{equation}\label{eq:error}
    E = \frac{d}{Max\_dist}
\end{equation}
where $d$ is the distance covered by the vehicle between two successive GPS data transmissions and $Max\_dist$ is the maximum distance that a vehicle can travel in that duration. Under normal condition, $E$ lies within $[0,1]$. If $E > 1$, then GPS spoofing attack is detected and the attack prevention phase is triggered.

We collected the roadmaps of Singapore using OpenStreetMap~\cite{openstreetmap} to evaluate our proposed countermeasure. We used the Traffic Control Interface in SUMO simulator~\cite{SUMO2018} to extract roadmap data from OpenStreetMap. We divided the roadmaps of Singapore into five regions - Central, East, West, North and South. For each region, we used the roadmap data to generate the data structures such as $Jn\_List$, $Rd\_List$, $Rd\_Jn$, $L$. We generated \textbf{380} random trips of cars on the roadmaps of Singapore with varying time duration of upto 10 minutes and collected the trajectory data of the cars from SUMO simulator. We implemented our proposed countermeasure in Python and ran our countermeasure on these generated trajectories. 
In our experiments, we considered that the onboard unit of the cars transmit the location data packet at every second. We used Equation~\eqref{eq:error} to detect the attack. 

Fig.~\ref{fig:error} shows the error function $E$ on running our proposed countermeasure on 380 random trips generated on the roadmaps of Singapore over a duration of 10 minutes. We observed that without GPS spoofing attack the value of $E$ remains below $1$ for all the trajectories of the vehicle. We observed that $E$ varies between $0$ and $1$ depending on the distance the vehicle has covered at different time of its journey. However, without GPS spoofing attack, $E$ always lies between $0$ and $1$. Fig.~\ref{fig:attackseries} shows $5$ random trajectories of vehicles from different regions of Singapore without GPS spoofing attack (in solid lines) and with GPS spoofing attack (in dashed lines). We observed that on launching GPS spoofing attack, the value of $E$ immediately exceeds $1$ for all the trajectories, \emph{i.e.}, line~8 of Algorithm~\ref{algo:attackdetect} gets satisfied for all the trajectories and $AttackPrevent()$ algorithm is called to authenticate $t$ subsequent message transmissions. In our experiments we considered $t$ to be $10$, however, $t$ can vary depending on the load on the network, number of operational cars in the fleet, frequency of spoofing attacks, etc. We used Secure Hashed Algorithm (SHA-512) on the vehicle ID (message) and the secure key shared between the vehicle and the backend server during the pre-processing phase as the key to generate the message digest for authenticating subsequent data transmissions~\cite{sha512}. We observed that the average time required to generate the message digest is approximately \textbf{0.3 ms} which is negligibly small. We also observed that the data packet size increases from \textbf{120 bytes} to \textbf{250 bytes} on incorporating the message digest along with the message. However, IEEE 802.11p allows upto 350 bytes of data in one transmission~\cite{ieee80211p}. Hence, the number of transmissions remain the same even after incorporating the attack prevention steps.

\begin{figure}
    \centering
    \includegraphics[height=4cm]{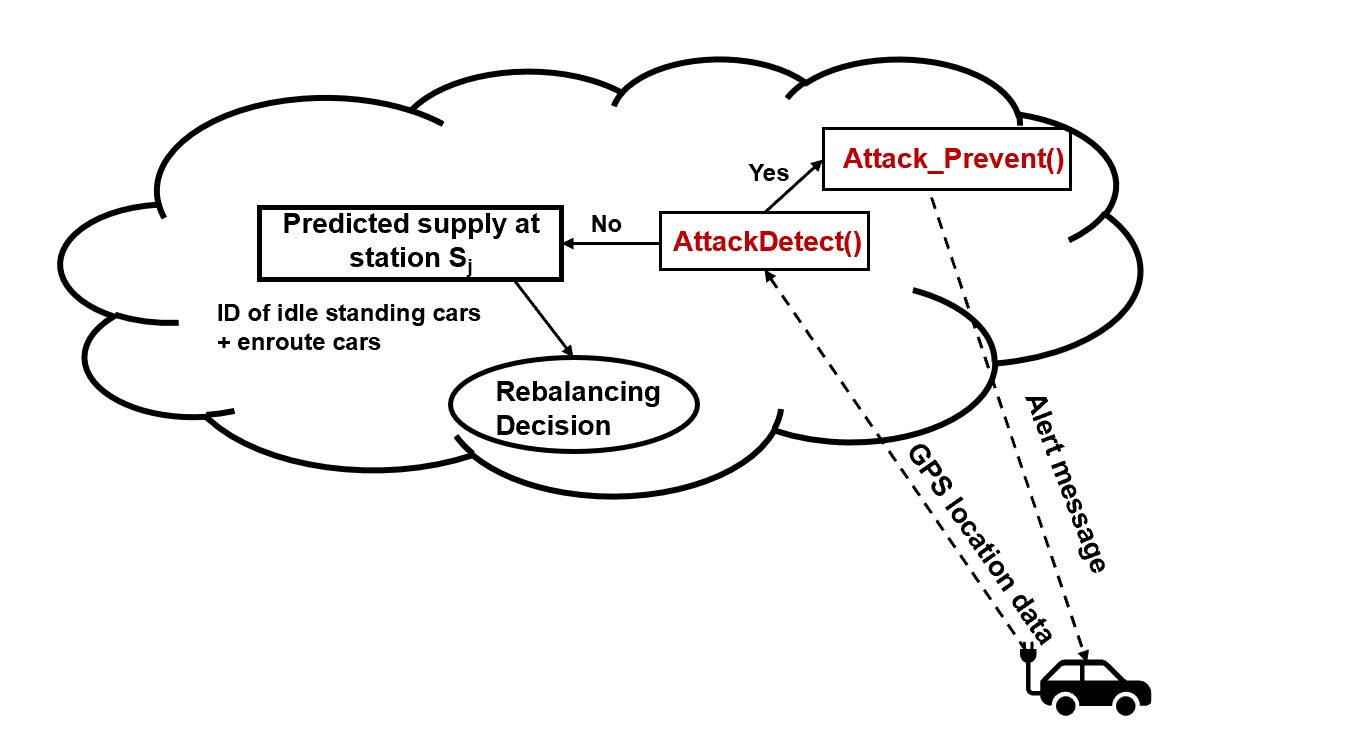}
    \caption{Integration of the countermeasure at the backend server}
    \label{fig:integration}
\end{figure}

\vspace{0.5em}\noindent
\textbf{Integration of the countermeasure :} We propose to integrate the attack detection algorithm  ($AttackDetect()$) at the backend server before the vehicle supply is predicted. This helps us to verify the estimated location of the vehicle before predicting the vehicle supply for the charging stations in the network. If an attack is detected, then the attack prevention algorithm ($AttackPrevent()$) is triggered at the backend server and an alert message is sent to the respective vehicle so that the subsequent $t$ message transmissions are associated with HMAC for authentication. Fig.~\ref{fig:integration} shows a schematic diagram of the integration of the countermeasure in the backend server.

\section{Conclusion}\label{sec:conclusion}
\noindent
In this work, we presented GPS spoofing attack in vehicle rebalancing application. To detect and prevent the attack, we proposed a location tracking technique that can validate the current location of the vehicle based on its previous location and road maps. We ran our experiments on the road maps of Singapore and observed that our technique was able to detect GPS spoofing attack immediately under all conditions. However, in this work, we considered a simple system model without any traffic in the system. We want to address the same problem for a more complex system model with traffic data into consideration.

\section{Acknowledgement}\label{sec:ack}
\noindent
This study is supported under the RIE2020 Industry Alignment Fund - Industry Collaboration Projects (IAF-ICP) Funding Initiative, as well as cash and in-kind contributions from the industry partner(s).

\bibliographystyle{plain}
\scriptsize
\bibliography{ref}

\begin{thebibliography}{10}

\bibitem{cyberattack2021}
Emad Aliwa, Omer Rana, Charith Perera, and Peter Burnap.
\newblock Cyberattacks and countermeasures for in-vehicle networks.
\newblock {\em ACM Comput. Surv.}, 54(1), mar 2021.

\bibitem{a2020arena}
Fabio Arena, Giovanni Pau, and Alessandro Severino.
\newblock A review on ieee 802.11p for intelligent transportation systems.
\newblock {\em Journal of Sensor and Actuator Networks}, 9(2), 2020.

\bibitem{ieee80211p}
Alessandro Bazzi, Stefania Bartoletti, Alberto Zanella, and Vincent Martinez.
\newblock Performance analysis of ieee 802.11p preamble insertion in c-v2x
  sidelink signals for co-channel coexistence.
\newblock {\em Vehicular Communications}, 45:100710, 2024.

\bibitem{sha512}
Himanshu~N. Bhonge, Monish~K. Ambat, and B.~R. Chandavarkar.
\newblock An experimental evaluation of sha-512 for different modes of
  operation.
\newblock In {\em 2020 11th International Conference on Computing,
  Communication and Networking Technologies (ICCCNT)}, pages 1--6, 2020.

\bibitem{brar2020}
Avalpreet~Singh Brar and Rong Su.
\newblock Ensuring service fairness in taxi fleet management.
\newblock In {\em 2020 IEEE 23rd International Conference on Intelligent
  Transportation Systems (ITSC)}, pages 1--6, 2020.

\bibitem{comprehensive2011}
Stephen Checkoway, Damon McCoy, Brian Kantor, Danny Anderson, Hovav Shacham,
  Stefan Savage, Karl Koscher, Alexei Czeskis, Franziska Roesner, and Tadayoshi
  Kohno.
\newblock Comprehensive experimental analyses of automotive attack surfaces.
\newblock In {\em Proceedings of the 20th USENIX Conference on Security},
  SEC'11, page~6, USA, 2011. USENIX Association.

\bibitem{demissie2022}
Merkebe Demissie and Lina Kattan.
\newblock Estimation of truck origin-destination flows using gps data.
\newblock {\em Transportation Research Part E Logistics and Transportation
  Review}, 159, 02 2022.

\bibitem{cybersecurity2020}
Zeinab El-Rewini, Karthikeyan Sadatsharan, Daisy~Flora Selvaraj, Siby~Jose
  Plathottam, and Prakash Ranganathan.
\newblock Cybersecurity challenges in vehicular communications.
\newblock {\em Vehicular Communications}, 23:100214, 2020.

\bibitem{wbss}
Andreas Festag.
\newblock Standards for vehicular communication—from ieee 802.11p to 5g.
\newblock {\em e \& i Elektrotechnik und Informationstechnik}, 132, 09 2015.

\bibitem{Deeppose2022}
Peng Jiang, Hongyi Wu, and Chunsheng Xin.
\newblock Deeppose: Detecting gps spoofing attack via deep recurrent neural
  network.
\newblock {\em Digit. Commun. Networks}, 8:791--803, 2022.

\bibitem{gpsspoofing2021}
Mohsin Kamal, Arnab Barua, Christian Vitale, Christos Laoudias, and Georgios
  Ellinas.
\newblock Gps location spoofing attack detection for enhancing the security of
  autonomous vehicles.
\newblock In {\em 2021 IEEE 94th Vehicular Technology Conference
  (VTC2021-Fall)}, pages 1--7, 2021.

\bibitem{inlinespoofing2010}
B.M. Ledvina, William Bencze, B.~Galusha, and I.~Miller.
\newblock An in-line anti-spoofing device for legacy civil gps receivers.
\newblock {\em Proceedings of the 2010 International Technical Meeting of the
  Institute of Navigation}, pages 698--712, 01 2010.

\bibitem{SUMO2018}
Pablo~Alvarez Lopez, Michael Behrisch, Laura Bieker-Walz, Jakob Erdmann,
  Yun-Pang Fl{\"o}tter{\"o}d, Robert Hilbrich, Leonhard L{\"u}cken, Johannes
  Rummel, Peter Wagner, and Evamarie Wie{\ss}ner.
\newblock Microscopic traffic simulation using sumo.
\newblock In {\em The 21st IEEE International Conference on Intelligent
  Transportation Systems}. IEEE, 2018.
\newblock \url{https://elib.dlr.de/124092/}.

\bibitem{miao2019}
Fei Miao, Shuo Han, Shan Lin, Qian Wang, John~A. Stankovic, Abdeltawab Hendawi,
  Desheng Zhang, Tian He, and George~J. Pappas.
\newblock Data-driven robust taxi dispatch under demand uncertainties.
\newblock {\em IEEE Transactions on Control Systems Technology},
  27(1):175--191, 2019.

\bibitem{gpsspoofing2023}
Pavlo Mykytyn, Marcin Brzozowski, Zoya Dyka, and Peter Langendoerfer.
\newblock Gps-spoofing attack detection mechanism for uav swarms.
\newblock In {\em 2023 12th Mediterranean Conference on Embedded Computing
  (MECO)}, pages 1--8, 2023.

\bibitem{openstreetmap}
{OpenStreetMap}.
\newblock \url{https://www.openstreetmap.org/#map=15/36.1514/-86.7876}.

\bibitem{marco2012}
Marco Pavone, Stephen~L Smith, Emilio Frazzoli, and Daniela Rus.
\newblock Robotic load balancing for mobility-on-demand systems.
\newblock {\em The International Journal of Robotics Research}, 31(7):839--854,
  2012.

\bibitem{directional}
Ram Ramanathan, Jason Redi, Cesar Santivanez, David Wiggins, and Stephen Polit.
\newblock Ad hoc networking with directional antennas: A complete system
  solution.
\newblock {\em Selected Areas in Communications, IEEE Journal on}, 23:496 --
  506, 04 2005.

\bibitem{gpsspoof2012}
Zhenghao Zhang, Matthew Trinkle, Lijun Qian, and Husheng Li.
\newblock Quickest detection of gps spoofing attack.
\newblock In {\em MILCOM 2012 - 2012 IEEE Military Communications Conference},
  pages 1--6, 2012.

\end{thebibliography}
\end{document}